\documentclass[10pt]{iopart}
\usepackage{graphicx}
\eqnobysec

\newcommand{\beq}{\begin{equation}}
\newcommand{\eeq}{\end{equation}}
\newcommand{\beqa}{\begin{eqnarray}}
\newcommand{\eeqa}{\end{eqnarray}}

\newcommand{\dd}{{\rm d}}
\newcommand{\eps}{{\varepsilon}}
\newcommand{\frad}[2]{{\displaystyle{\displaystyle#1\over\displaystyle#2}}}
\newcommand{\fras}[2]{{\scriptstyle{\scriptstyle#1\over\scriptstyle#2}}}
\newcommand{\half}{\fras{1}{2}}
\newcommand{\lam}{\lambda}
\renewcommand{\max}{{\rm max}}
\newcommand{\mean}[1]{\langle#1\rangle}
\renewcommand{\min}{{\rm min}}
\newcommand{\ra}{{\rm RA}}
\newcommand{\stackunder}[2]{\mathrel{\mathop{#1}\limits_{#2}}}
\renewcommand{\u}{{\bullet}}
\newcommand{\z}{{\circ}}
\newcommand{\B}{{\cal B}}
\newcommand{\C}{{\cal C}}
\newcommand{\I}{{\cal I}}
\newcommand{\J}{{\cal J}}
\newcommand{\N}{{\cal N}}
\newcommand{\T}{{\bf T}}

\begin{document}

\title[Configurational entropy in one dimension]
{A renewal approach to configurational entropy in~one dimension}

\author{P L Krapivsky$^{1,2}$ and J M Luck$^3$}

\address{$^1$ Department of Physics, Boston University, Boston, MA 02215, USA}

\address{$^2$ Santa Fe Institute, Santa Fe, NM 87501, USA}

\address{$^3$ Universit\'e Paris-Saclay, CNRS, CEA, Institut de Physique Th\'eorique,
91191~Gif-sur-Yvette, France}

\begin{abstract}
We introduce a novel approach,
inspired from the theory of renewal processes,
to determine the configurational entropy
of ensembles of constrained configurations of particles on the one-dimensional lattice.
The proposed method can deal with all local rules involving only the lengths of clusters
of occupied and empty sites.
Within this scope, this method is both more systematic
and easier to implement than the transfer-matrix approach.
It is illustrated in detail on the $k$-mer deposition model
and on ensembles of trapped Rydberg atoms with blockade range $b$.
\end{abstract}

\eads{\mailto{pkrapivsky@gmail.com},\mailto{jean-marc.luck@ipht.fr}}

\maketitle

\section{Introduction}
\label{intro}

A variety of complex systems,
including spin glasses, structural glasses and granular matter,
are known to possess many metastable states at low enough temperature
and/or high enough density~\cite{thou,kirk,mpv,gotze,bm,debe,bb}.
These metastable states are usually defined by dynamical considerations.
They have been given various names in different contexts,
including valleys, pure states, quasi-states and inherent structures.
The number $\N(N,E)$ of metastable states with energy density $E$
usually grows exponentially with the system size $N$, as
\beq
\N(N,E)\sim\exp(N\,S(E)),
\label{sconf}
\eeq
where $S(E)$ is the configurational entropy, or complexity, of the system~\cite{jackle,palmer}.

The one-dimensional geometry is a convenient playground
to investigate the statistics of these metastable states.
There, metastability only takes place at zero temperature,
and valleys typically consist of single blocked configurations.
A broad range of one-dimensional kinetic spin models
possess an exponentially large number of blocked configurations,
which identify with the attractors of zero-temperature dynamics.
This class includes pristine models,
such as the ferromagnetic Ising chain with Kawasaki dynamics~\cite{cks,dsk},
disordered models, such as the Ising spin glass with single-spin-flip dynamics~\cite{dg,msj},
and a class of kinetically constrained spin models~\cite{pfa,pje,pse,crrs,dl1,dl2,dl3,pb,ds}
and of related lattice gas models~\cite{pf,ef,kpri,klin,kkra}.
The zero-temperature dynamics of these kinetically constrained models
is fully irreversible.
In most cases, it can be mapped onto RSA (random sequential adsorption)
or CSA (cooperative sequential adsorption) models,
where objects are deposited sequentially
on an initially empty substrate~\cite{evans,talbot,kinetic}.
This mapping brings several simplifications.
Blocked configurations often have a local characterization
in terms of forbidden patterns.
The configurational entropy counting the latter configurations
can be determined either by a direct combinatorial reasoning~\cite{crrs,d1,ld,dos1,dos2,kld}
or by using the transfer-matrix approach~\cite{dl3,ds}.
The dynamics of this class of models can also be solved exactly by an analytical approach
based on considering empty intervals
(see~\cite{gl,epjst} for overviews).
These models are strongly non-ergodic.
The energy density~$E_\infty$ at the end of the zero-temperature dynamical process,
where the system is blocked in one of its many attractors,
depends continuously on the energy density $E_0$ of the initial state.
For the most probable infinite-temperature initial state ($E_0=0$),
the final energy density $E_\infty$ is slightly different from
the most probable one~$E_\star$, which maximizes the configurational entropy $S(E)$.
This testifies a weak violation of the flatness hypothesis
originally formulated by Edwards for slowly compacting granular matter
(see~\cite{edrev} for a recent review).

The physics of ultracold atoms provides another motivation
for investigating the configurational entropy $S(\rho)$
of statistical ensembles of particles defined by local constraints,
where $\rho$ now denotes the particle density.
Trapped Rydberg atoms provide a promising experimental playground
for quantum information processing, quantum computation and quantum simulation~\cite{swm}.
The strong interactions between Rydberg atoms generate a blockade,
preventing the excitation of further
Rydberg atoms in some vicinity of an already existing one~\cite{jcz,lbr,pdl,vhb,hgs,bsk}.
Under certain conditions, the Rydberg blockade yields patterns of atoms,
either on periodic optical lattices or on more complex graphs and networks,
which are somehow similar to the blocked configurations of RSA models~\cite{san}.
A simple setting is that of a one-dimensional optical lattice,
where each lattice site occupied by a Rydberg atom
must have at least~$b$ empty sites on either side~\cite{kld}.
The integer $b\ge1$ is referred to as the blockade range of the model.
The problem with blockade range $b=1$ on more complex graphs
amounts to counting the `maximal independent sets' on the underlying graph.
This combinatorial task is known to be a hard NP-complete one in general~\cite{lawler}.
Quite recently, experimental systems of Rydberg atoms have been demonstrated
to be able to find maximal independent sets in some geometries~\cite{ebadi,qo2}.

The present paper has been elaborated concomitantly with the work
by Do\v{s}li\'c et~al.~\cite{crew}.
Here, our main goal is to introduce a novel exact method to determine
the configurational entropy $S(\rho)$
of statistical ensembles of constrained configurations of particles
on the one-dimensional lattice
defined by local rules.
This approach, inspired from the theory of renewal processes,
provides an alternative to the usual methods recalled above,
based either on direct combinatorial reasoning or on the transfer-matrix approach.
The usage of the renewal approach is however limited to constraints involving
the lengths of clusters of occupied and empty sites.
Within this scope, it is more systematic and easier to implement than both other methods.
The renewal approach itself is described in section~\ref{renewal}.
Section~\ref{thermo} is devoted to the analysis
of quantities pertaining to the thermodynamic limit,
including the cumulants of the number of particles and the configurational entropy.
Fully explicit expressions are presented in section~\ref{examples}
for three simple statistical ensembles.
The main focus is then on the $k$-mer deposition model
and on assemblies of trapped Rydberg atoms with blockade range~$b$.
Both models are essentially equivalent to each other, with $k=b+1$.
They are respectively investigated in sections~\ref{kmer} and~\ref{rydberg}.
A brief discussion is given in section~\ref{disc}.
Two appendices are respectively devoted to an extension of the renewal approach
to several species of particles,
and to a reminder on the transfer-matrix formalism.

\section{Renewal approach}
\label{renewal}

Within the present approach,
a configuration $\C$ is a sequence of particles (i.e., occupied sites, noted $\u$)
and holes (i.e., empty sites, noted $\z$) on the half-infinite one-dimensional chain.
The basic variables are the lengths $i_1$, $j_1$, $i_2$, $j_2,$ $\dots$
of the clusters of contiguous occupied and empty sites.
The configuration thus reads
\beq
\C=\underbrace{\u\cdots\u}_{i_1}
\underbrace{\z\cdots\z}_{j_1}
\underbrace{\u\cdots\u}_{i_2}
\underbrace{\z\cdots\z}_{j_2}\cdots
\label{init1}
\eeq
if the first site is occupied,
and
\beq
\C=\underbrace{\z\cdots\z}_{j_1}
\underbrace{\u\cdots\u}_{i_1}
\underbrace{\z\cdots\z}_{j_2}
\underbrace{\u\cdots\u}_{i_2}\cdots
\label{init0}
\eeq
if the first site is empty.

We define a statistical ensemble of configurations
by putting the following constraints on the (mutually independent) cluster lengths:

\begin{itemize}

\item
the lengths $i_1$, $i_2$, $\dots$ of clusters of occupied sites
belong to some set $\I$ of integers.

\item
the lengths $j_1$, $j_2$, $\dots$ of clusters of empty sites
belong to some set $\J$ of integers.

\end{itemize}

The first quantity of interest is
the total number $\N_N$ of configurations $\C_N$ on a finite system of length $N$.
It is understood that the rightmost cluster ends exactly at site $N$.
In other words, at variance with the usual approach to renewal processes in continuous time,
no overhang is permitted.

We have
\beqa
\N_N
&=&\sum_{i_1\in\I}\delta_{i_1,N}+\sum_{i_1\in\I}\sum_{j_1\in\J}\delta_{i_1+j_1,N}
\nonumber\\
&&+\sum_{i_1\in\I}\sum_{j_1\in\J}\sum_{i_2\in\I}\delta_{i_1+j_1+i_2,N}+\cdots
\nonumber\\
&+&\sum_{j_1\in\J}\delta_{j_1,N}+\sum_{j_1\in\J}\sum_{i_1\in\I}\delta_{j_1+i_1,N}
\nonumber\\
&&+\sum_{j_1\in\J}\sum_{i_1\in\I}\sum_{j_2\in\J}\delta_{j_1+i_1+j_2,N}+\cdots,
\eeqa
where the first two and last two lines
respectively correspond to~(\ref{init1}) and~(\ref{init0}).

An efficient way of resumming the above series
is suggested by a formal analogy with
the theory of renewal processes~\cite{cox,cm,feller}
(see~\cite{glrenew,sbm} for presentations by physicists).
Introduce the generating series
\beq
I(z)=\sum_{i\in\I}z^i,
\qquad
J(z)=\sum_{j\in\J}z^j,
\eeq
associated with the sets $\I$ and $\J$ encoding the definition of the statistical ensemble.
The generating series of the numbers of configurations,
\beq
\N(z)=\sum_{N\ge0}\N_Nz^N,
\eeq
then reads
\beqa
\N(z)=1+\N_I(z)+\N_J(z),
\nonumber\\
\N_I(z)=I(z)+I(z)J(z)+I(z)J(z)I(z)+\cdots,
\nonumber\\
\N_J(z)=J(z)+J(z)I(z)+J(z)I(z)J(z)+\cdots
\label{nser}
\eeqa
The initial term $\N_0=1$ is somehow conventional,
whereas $\N_I(z)$ and $\N_J(z)$
are in correspondence with~(\ref{init1}) and~(\ref{init0}).
These series obey the `renewal equations'
\beq
\N_I(z)=I(z)(1+\N_J(z)),\qquad
\N_J(z)=J(z)(1+\N_I(z)).
\eeq
Solving the above linear equations,
we obtain our first key result:
\beq
\N(z)=\frac{(1+I(z))(1+J(z))}{1-I(z)J(z)}.
\label{nres}
\eeq

The second quantity of interest is
the number $\N_{N,M}$ of configurations on a system of length $N$,
comprising exactly $M$ particles,
i.e., $M$ occupied sites and $N-M$ empty ones.
This quantity is advantageously encoded in the partition function
\beq
Z_N(x)=\sum_{M=0}^N\N_{N,M}x^M,
\label{zdef}
\eeq
obtained by attributing a positive weight $x$ to each particle.
We have therefore
\beq
\mean{x^M}=\frac{Z_N(x)}{Z_N(1)},
\label{avexp}
\eeq
where brackets denote an average over all configurations $\C_N$ of length $N$.

The partition function reads alternatively
\beq
Z_N(x)=\sum_{\C_N}x^{M(\C_N)},
\eeq
where the sum runs over all configurations $\C_N$, with
\beq
M(\C_N)=i_1+i_2+\cdots
\eeq
being the number of particles in the configuration $\C_N$.
In analogy with~(\ref{nser}),
the generating series of the partition functions, namely
\beq
Z(z,x)=\sum_{N\ge0}Z_N(x)z^N=\sum_{N\ge0}\sum_{M=0}^N\N_{N,M}z^Nx^M,
\label{hzdef}
\eeq
reads
\beqa
Z(z,x)=1
&+&I(xz)+I(xz)J(z)+I(xz)J(z)I(xz)+\cdots
\nonumber\\
&+&J(z)+J(z)I(xz)+J(z)I(xz)J(z)+\cdots
\eeqa
We thus obtain our second key result:
\beq
Z(z,x)=\frac{(1+I(xz))(1+J(z))}{1-I(xz)J(z)}.
\label{zres}
\eeq
For $x=1$, where all configurations have the same weight, we have
\beq
Z(z,1)=\N(z),
\label{un}
\eeq
and therefore
\beq
Z_N(1)=\N_N=\sum_{M=0}^N\N_{N,M},
\eeq
as should be.

The expressions~(\ref{nres}) and~(\ref{zres}) hold in full generality.
They are the most useful in the rational case
where both generating series
\beq
I(z)=\frac{A_I(z)}{B_I(z)},\qquad
J(z)=\frac{A_J(z)}{B_J(z)}
\eeq
are rational functions of $z$.
The expressions~(\ref{nres}) and~(\ref{zres}) then read
\beq
\N(z)=\frac{C(z)}{D(z)},
\label{ncd}
\eeq
where
\beqa
C(z)=(A_I(z)+B_I(z))(A_J(z)+B_J(z)),
\nonumber\\
D(z)=B_I(z)B_J(z)-A_I(z)A_J(z)
\eeqa
are polynomials in $z$,
and
\beq
Z(z,x)=\frac{C(z,x)}{D(z,x)},
\label{zcd}
\eeq
where
\beqa
C(z,x)=(A_I(xz)+B_I(xz))(A_J(z)+B_J(z)),
\nonumber\\
D(z,x)=B_I(xz)B_J(z)-A_I(xz)A_J(z)
\eeqa
are polynomials in $z$ and $x$.
We have consistently $C(z)=C(z,1)$ and $D(z)=D(z,1)$.

The complexity of an ensemble can be measured by
the degree $\Delta$ of the polynomial~$D(z)$,
which coincides with the degree in $z$ of $D(z,x)$.
In particular, as a consequence of~(\ref{ncd}),
the total numbers $\N_N$ of configurations obey a linear recursion
with constant integer coefficients,
whose number of terms is at most $\Delta+1$.
Explicit examples will be given in~(\ref{nfibo}) and~(\ref{npado}).

The rational class encompasses a panoply of ensembles.
If $\I$ is a finite set,~$I(z)$ is a polynomial in $z$.
If $\I=\{b,2b,3b,\dots\}$ is the sublattice of index $b$, we have
\beq
I(z)=\frac{z^b}{1-z^b}.
\eeq
The rational class therefore comprises all cases where the sets $\I$ and $\J$
are either finite sets, sublattices, or any finite unions of such sets.
From now on, we restrict ourselves to this rational class of ensembles.

\section{Thermodynamic limit}
\label{thermo}

The above approach
provides an efficient framework to investigate
the distribution of the number $M$ of particles
in the thermodynamic limit of very large systems.

In the simplest cases,
the renewal approach also gives access to the full finite-size combinatorics
encoded in the numbers $\N_{N,M}$ (see section~\ref{examples}).

\subsection{Cumulants of number of particles}
\label{cum}

A first side of the problem concerns the cumulants
of the number $M$ of particles in a very large system of size $N$.

The rational expression~(\ref{zcd}) of the generating series $Z(z,x)$
implies an asymp\-totic exponential behavior
of the partition function $Z_N(x)$ at large $N$, namely
\beq
Z_N(x)\sim z_0(x)^{-N},
\label{zasy}
\eeq
where $z_0(x)$ is the nearest zero of the polynomial $D(z,x)$,
i.e., that with smallest modulus among the $\Delta$ zeros.
Since the weight $x$ is real, the partition functions $Z_N(x)$ are positive,
and so $z_0(x)$ is real and positive.
Depending on $x$, $z_0(x)$ may be either smaller or larger than unity.

Setting
\beq
x=\e^\beta,
\label{xbeta}
\eeq
where $\beta$ is a fictitious inverse temperature,
being either positive or negative,
the identity~(\ref{avexp}) yields
\beq
\mean{\e^{\beta M}}\sim\e^{NF(\beta)},
\label{egm}
\eeq
where the configurational free energy $F(\beta)$ reads
\beq
F(\beta)=\ln z_\star-\ln z_0(\e^\beta),\qquad
z_\star=z_0(1).
\label{fdef}
\eeq
We assume that the problem is non-trivial,
in the sense that at least one of the sets~$\I$ and~$\J$ contains more than one point.
The free energy then has a full `high-temperature' power-series expansion of the form
\beq
F(\beta)=\sum_{n\ge1}\frac{c_n}{n!}\,\beta^n.
\label{fexpan}
\eeq
As a consequence of~(\ref{egm}),
all cumulants of $M$ grow linearly with the system size~$N$,~as
\beq
\mean{M^n}_c\approx c_nN.
\eeq
In particular, the mean and the variance of the particle number grow as
\beqa
\mean{M}\approx c_1N,
\nonumber\\
\mean{M^2}_c=\mean{M^2}-\mean{M}^2\approx c_2N.
\label{c1c2}
\eeqa
The first cumulant amplitude $c_1$ is nothing but the most probable density $\rho_\star$
(see~(\ref{rcid})), as should be.

\subsection{Configurational entropy}
\label{sent}

Let us turn to the asymptotic growth of the number $\N_{N,M}$
of configurations $\C_N$ of a system of size $N$
comprising $M$ particles, i.e., $M$ occupied sites.
This number is expected on general grounds to scale as (see~(\ref{sconf}))
\beq
\N_{N,M}\sim\exp(N\,S(\rho)),
\label{smicro}
\eeq
where the microcanonical configurational entropy $S(\rho)$ is a function of
the particle density
\beq
\rho=\frac{M}{N}.
\eeq

The total number $\N_N$ of configurations $\C_N$
on a system of size $N$ therefore also grows exponentially, as
\beq
\N_N\sim\e^{NS_\star},\qquad
S_\star=\,\stackunder{\max}{\rho}S(\rho).
\eeq
The value $\rho_\star$ of the density
where the microcanonical entropy $S(\rho)$ reaches its maximum $S_\star$
represents the most probable density in the thermodynamic limit.

The configurational entropy $S(\rho)$ can be derived as follows.
The partition function $Z_N(x)$ can be estimated
by using~(\ref{zdef}),~(\ref{zasy}) and~(\ref{smicro}).
We thus obtain
\beq
Z_N(x)\sim\int\e^{N[S(\rho)+\rho\ln x]}\,\dd\rho\sim\e^{-N\ln z_0(x)}.
\eeq
The saddle-point method yields
\beq
\ln z_0(x)=-\stackunder{\max}{\rho}[S(\rho)+\rho\ln x].
\eeq
This equality expresses that the functions $S(\rho)$ and $\ln z_0(x)$
are Legendre transforms of each other.
We have therefore
\beq
S(\rho)=-\ln z_0(x)-\rho\ln x,
\label{sres}
\eeq
with
\beq
\rho=-\frac{\dd\ln z_0}{\dd\ln x}=-\frac{x}{z_0}\,\frac{\dd z_0}{\dd x}
=\left(\frac{x\,\partial D/\partial x}{z\,\partial D/\partial z}\right)_{z=z_0(x)}.
\label{rhores}
\eeq
We have alternatively, using~(\ref{xbeta}) and~(\ref{fdef}),
\beq
S(\rho)=-\ln z_\star+F(\beta)-\beta\rho,\qquad
\rho=\frac{\dd F}{\dd\beta}.
\label{rhof}
\eeq

The maximal entropy $S_\star$ is reached for $x=1$, i.e., $\beta=0$,
as should be, since this is the case where all configurations $\C_N$ have equal weights.
We have
\beq
S_\star=-\ln z_\star.
\label{szstar}
\eeq
The series expansion~(\ref{fexpan})
ensures that the corresponding density reads
\beq
\rho_\star=c_1,
\label{rcid}
\eeq
and that $S(\rho)$ departs quadratically from its maximum~$S_\star$, as
\beq
S(\rho)\approx S_\star-\frac{(\rho-\rho_\star)^2}{2c_2}.
\label{squadra}
\eeq

Another interpretation of the configurational entropy $S(\rho)$ is as follows.
The probability $P_N(\rho)$ of observing
a configuration $\C_N$ with an atypical density $\rho\ne\rho_\star$
is given by a large-deviation formula of the form
\beq
P_N(\rho)\sim\exp(-N\Sigma(\rho)),
\eeq
where the large-deviation function reads
\beq
\Sigma(\rho)=S_\star-S(\rho).
\label{sigap}
\eeq
Using~(\ref{rhof}) and~(\ref{szstar}), we have alternatively
\beq
\Sigma(\rho)=\beta\rho-F(\beta).
\label{sigr}
\eeq
The large-deviation function is positive,
and vanishes quadratically in the vicinity of~$\rho_\star$, according to
(see~(\ref{squadra}))
\beq
\Sigma(\rho)\approx\frac{(\rho-\rho_\star)^2}{2c_2}.
\eeq

\section{Three simple cases}
\label{examples}

The complexity of a statistical ensemble has been argued to be dictated
by the degree~$\Delta$ of the polynomial $D(z)$.
In this section we illustrate the above general results
on the only three examples (up to symmetries)
where either $\Delta=1$ (Section~\ref{flat})
or $\Delta=2$ (Sections~\ref{isola} and~\ref{even}).
These examples essentially exhaust all cases where fully explicit expressions can be obtained,
either for finite systems or in the thermodynamic limit.

\subsection{Flat ensemble}
\label{flat}

The first and simplest ensemble is the flat one,
where all cluster lengths are equally permitted.
Equivalently, each site of the lattice is independently either occupied or empty.
We have $\I=\J=\{1,2,3,\dots\}$, so that
\beq
I(z)=J(z)=\frac{z}{1-z}.
\eeq
The formula~(\ref{nres}) reads
\beq
\N(z)=\frac{1}{1-2z},
\eeq
so that
\beq
\N_N=2^N.
\label{flatn}
\eeq
The formula~(\ref{zres}) reads
\beq
Z(z,x)=\frac{1}{1-(1+x)z}.
\label{z1}
\eeq
This is the only instance where the polynomials
\beq
D(z)=1-2z,\qquad
D(z,x)=1-(1+x)z
\eeq
have degree $\Delta=1$ in $z$.
The expression~(\ref{z1}) yields
\beq
\N_{N,M}={N\choose M},
\label{flatnm}
\eeq
where $M$ ranges from $0$ to $N$.

Asymptotic expressions in the thermodynamic limit are also simple.
We have
\beq
z_0(x)=\frac{1}{1+x},
\eeq
so that~(\ref{fdef}) yields
\beq
F(\beta)=\ln\frac{1+\e^\beta}{2}=\frac{\beta}{2}+\ln\cosh\frac{\beta}{2}.
\eeq
We have therefore
\beq
\rho_\star=c_1=\frac{1}{2},
\eeq
in agreement with particle-hole symmetry,
whereas all higher-order odd cumulants vanish.
The even cumulant amplitudes are given by
\beq
c_n=\frac{2^n-1}{n}\,B_n,
\eeq
where $B_n$ are the Bernoulli numbers, i.e.,
\beq
c_2=\frac{1}{4},\quad
c_4=-\frac{1}{8},\quad
c_6=\frac{1}{4},\quad
c_8=-\frac{17}{16},\quad\hbox{etc.}
\eeq

The formula~(\ref{rhores}) implies that
$x$ and $z_0(x)$ can be expressed as rational functions of the density $\rho$:
\beq
x=\frac{\rho}{1-\rho},\qquad
z_0(x)=1-\rho,
\eeq
so that~(\ref{sres}) yields the explicit expression
\beq
S(\rho)=-\rho\ln\rho-(1-\rho)\ln(1-\rho)
\label{sent1}
\eeq
for $0<\rho<1$.
The above formula coincides with the well-known expression for the entropy of mixing.
It is the largest possible value for the configurational entropy $S(\rho)$
of an ensemble consisting of occupied and empty sites.
In particular,
\beq
S_\star=\ln 2\approx0.693147,
\eeq
in agreement with~(\ref{flatn}).
This is the largest possible value for the entropy $S_\star$.

\subsection{Isolated empty sites}
\label{isola}

This second ensemble is defined by the condition that empty sites are isolated.
The dual ensemble where occupied sites are isolated is simply obtained
from the present one by changing $M$ to $N-M$, and $\rho$ to $1-\rho$.
These two ensembles have already been considered in several contexts,
including the attractors of repulsion processes~\cite{rep}
and packings of disks in narrow channels~\cite{moo1,moo2}.

In the present case we have $\I=\{1,2,3,\dots\}$ and $\J=\{1\}$, so that
\beq
I(z)=\frac{z}{1-z},\qquad J(z)=z.
\eeq
The formula~(\ref{nres}) reads
\beq
\N(z)=\frac{1+z}{1-z-z^2},
\eeq
implying that the numbers $\N_N$ obey the recursion
\beq
\N_N=\N_{N-1}+\N_{N-2},
\label{nfibo}
\eeq
defining the Fibonacci numbers
\beq
F_n=\frac{1}{\sqrt{5}}
\left[\left(\frac{1+\sqrt{5}}{2}\right)^n-\left(\frac{1-\sqrt{5}}{2}\right)^n\right].
\label{fibo}
\eeq
These integers are listed in entry A000045 of the OEIS~\cite{OEIS},
together with many formulas and references.
The initial values $\N_0=1$, $\N_1=2$ lead to
\beq
\N_N=F_{N+2}.
\label{n2}
\eeq
The formula~(\ref{zres}) reads
\beq
Z(z,x)=\frac{1+z}{1-xz-xz^2}.
\label{z2}
\eeq
The polynomials
\beq
D(z)=1-z-z^2,\qquad
D(z,x)=1-xz-xz^2
\eeq
have degree $\Delta=2$ in $z$.
The expression~(\ref{z2}) yields after some algebra
\beq
\N_{N,M}={M+1\choose N-M}.
\eeq
The minimal value of $M$ is either $N/2$ (if $N$ is even) or $(N-1)/2$ (if $N$ is odd),
whereas its maximal value is $N$.

Asymptotic expressions in the thermodynamic limit are also simple.
We have
\beq
z_0(x)=\frac{\sqrt{x(x+4)}-x}{2x},
\eeq
so that~(\ref{fdef}) yields
\beq
F(\beta)=\beta+\ln\frac{1+\sqrt{1+4\e^{-\beta}}}{1+\sqrt{5}},
\eeq
and so
\beq
\rho_\star=c_1=\frac{5+\sqrt{5}}{10}\approx0.723606,
\eeq
and
\beq
c_2=\frac{\sqrt{5}}{25},\quad
c_3=\frac{\sqrt{5}}{125},\quad
c_4=-\frac{\sqrt{5}}{125},\quad\hbox{etc.}
\eeq

Even though the polynomial $D(z,x)$ is quadratic in $z$,
both $x$ and $z_0(x)$ can be expressed as rational functions of the density $\rho$:
\beq
x=\frac{(2\rho-1)^2}{\rho(1-\rho)},\qquad
z_0(x)=\frac{1-\rho}{2\rho-1},
\eeq
so that~(\ref{sres}) yields the explicit expression
\beq
S(\rho)=-(2\rho-1)\ln(2\rho-1)-(1-\rho)\ln(1-\rho)+\rho\ln\rho
\label{sent2}
\eeq
for $1/2<\rho<1$.
In particular,
\beq
S_\star=\ln\frac{1+\sqrt{5}}{2}\approx0.481211,
\label{sstar2}
\eeq
in agreement with~(\ref{n2}).

\subsection{Even particle clusters}
\label{even}

This third ensemble is defined by the condition that all clusters of particles
(i.e., of occupied sites) have even lengths.
We have $\I=\{2,4,6,\dots\}$ and $\J=\{1,2,3,\dots\}$, so~that
\beq
I(z)=\frac{z^2}{1-z^2},\qquad J(z)=\frac{z}{1-z}.
\eeq
The formula~(\ref{nres}) reads
\beq
\N(z)=\frac{1}{1-z-z^2},
\eeq
implying that the numbers of configurations are again given by Fibonacci numbers:
\beq
\N_N=F_{N+1}.
\label{n3}
\eeq
The formula~(\ref{zres}) reads
\beq
Z(z,x)=\frac{1}{1-z-x^2z^2}.
\label{z3}
\eeq
The polynomials
\beq
D(z)=1-z-z^2,\qquad
D(z,x)=1-z-x^2z^2
\eeq
have degree $\Delta=2$ in $z$.
The number $M$ of particles is necessarily even.
The expression~(\ref{z3}) yields after some algebra
\beq
\N_{N,M}={N-\half M\choose\half M}.
\eeq
The minimal value of $M$ is 0,
whereas its maximal value is either $N$ (if $N$ is even) or $N-1$ (if $N$ is odd).

Asymptotic expressions in the thermodynamic limit are also simple.
We have
\beq
z_0(x)=\frac{\sqrt{1+4x^2}-1}{2x^2},
\eeq
so that~(\ref{fdef}) yields
\beq
F(\beta)=\ln\frac{1+\sqrt{1+4\e^{2\beta}}}{1+\sqrt{5}},
\eeq
and so
\beq
\rho_\star=c_1=\frac{5-\sqrt{5}}{5}\approx0.552786,
\eeq
and
\beq
c_2=\frac{4\sqrt{5}}{25},\quad
c_3=-\frac{8\sqrt{5}}{125},\quad
c_4=-\frac{16\sqrt{5}}{125},\quad\hbox{etc.}
\eeq

Here again,
both $x^2$ and $z_0(x)$ can be expressed as rational functions of the density~$\rho$:
\beq
x^2=\frac{\rho(2-\rho)}{4(1-\rho)^2},\qquad
z_0(x)=\frac{2(1-\rho)}{2-\rho},
\eeq
so that~(\ref{sres}) yields the explicit expression
\beq
S(\rho)=\frac{1}{2}(2-\rho)\ln(2-\rho)-(1-\rho)\ln(2(1-\rho))-\frac{1}{2}\,\rho\ln\rho
\label{sent3}
\eeq
for $0<\rho<1$.
In particular,
\beq
S_\star=\ln\frac{1+\sqrt{5}}{2}\approx0.481211,
\eeq
in agreement with~(\ref{n3}),
is the same as in the previous ensemble (see~(\ref{sstar2})).

\begin{figure}
\begin{center}
\includegraphics[angle=0,width=0.6\linewidth,clip=true]{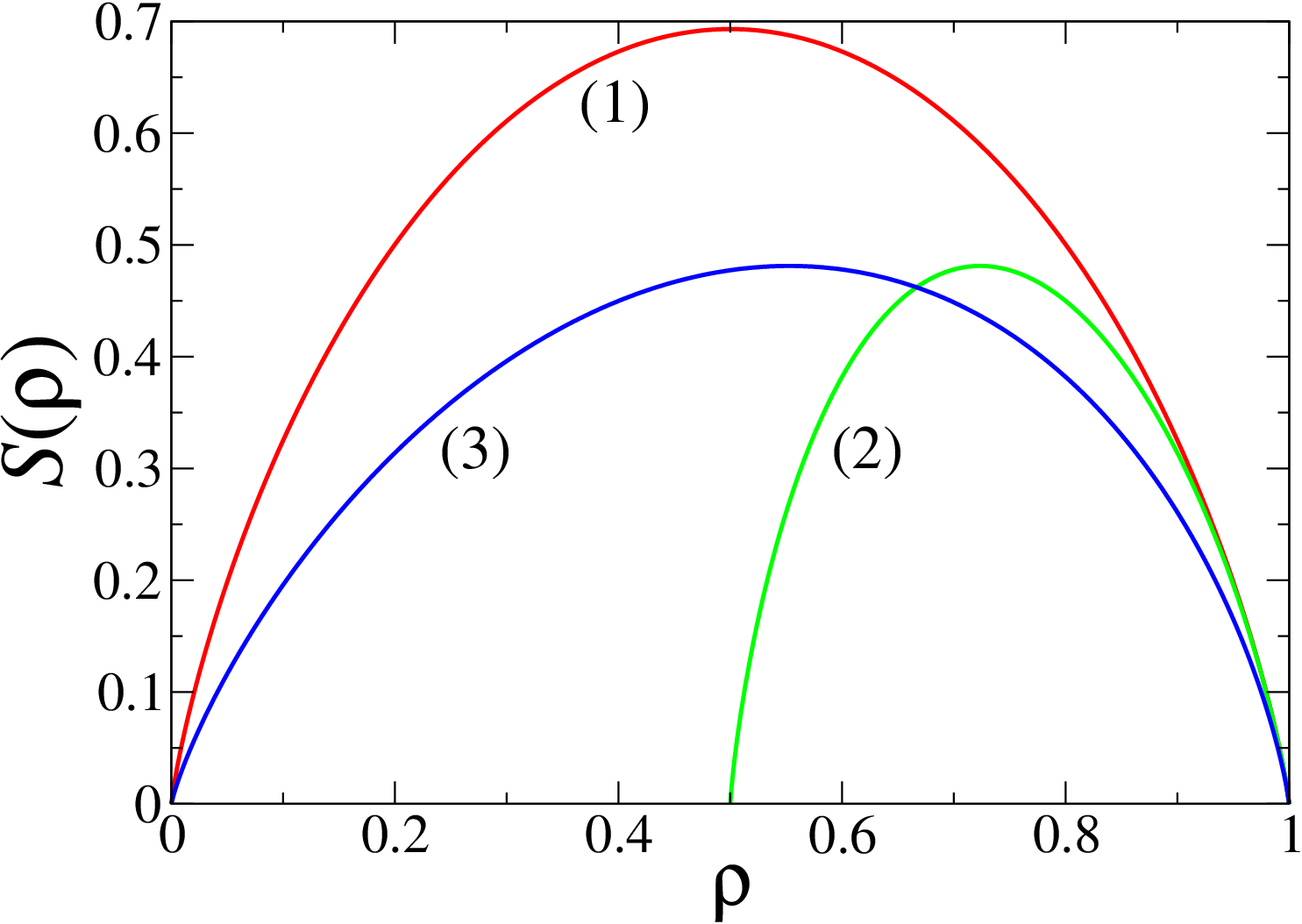}
\caption{\small
Configurational entropy $S(\rho)$
of the three simple ensembles investigated in section~\ref{examples},
against particle density $\rho$.
Red curve (1): flat ensemble (see~(\ref{sent1})).
Green curve (2): ensemble with isolated empty sites (see~(\ref{sent2})).
Blue curve (3): ensemble with even particle clusters (see~(\ref{sent3})).}
\label{exent}
\end{center}
\end{figure}

Figure~\ref{exent} shows plots of the configurational entropy $S(\rho)$
against the particle density $\rho$,
for the three simple ensembles investigated in section~\ref{examples}.

\section{$k$-mer deposition model}
\label{kmer}

\begin{figure}
\begin{center}
\includegraphics[angle=0,width=0.7\linewidth,clip=true]{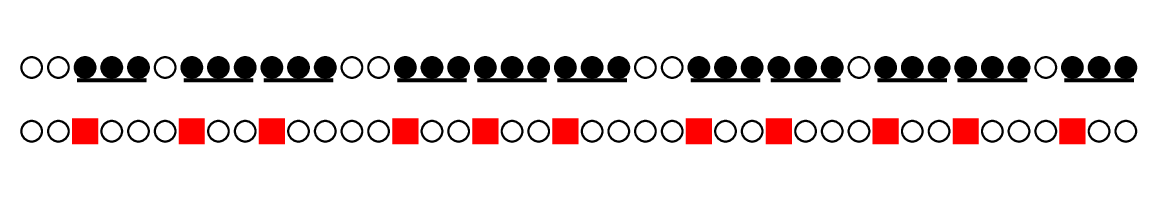}
\caption{\small
Top row:
typical blocked configuration of the trimer problem.
Bottom row:
corresponding configuration of Rydberg atoms with blockade range $b=2$
(see section~\ref{rydberg}).}
\label{seqs}
\end{center}
\end{figure}

In this section we investigate the problem of the deposition of $k$-mers,
i.e., clusters of~$k$ particles,
starting from an empty lattice~\cite{ghh,beg,bonnier}.
This is a prototypical example of the class of RSA models discussed
in the Introduction (see~\cite{evans,talbot,kinetic} for reviews).
The integer $k\ge2$ is the only parameter of the model,
with $k=2$ corresponding to dimers, $k=3$ to trimers, and so~on.
Our aim is to investigate the statistical ensemble of the blocked
(or jammed) configurations, where no single $k$-mer can be inserted any more.
These configurations consist of sequences of contiguous $k$-mers,
separated by clusters of empty sites whose length is at most $k-1$.
Figure~\ref{seqs} shows a typical blocked configuration of the trimer problem,
together with the corresponding configuration of Rydberg atoms with blockade range $b=2$
(see section~\ref{rydberg}).

The statistical ensemble of blocked configurations of $k$-mers
has been studied by combinatorial methods~\cite{dos1,dos2,kld}.
It is investigated by yet another approach in~\cite{crew}.

\subsection{General theory}

The ensemble of blocked configurations
can be described in terms of independent clusters,
and therefore studied by the renewal approach,
with $\I=\{k,2k,3k,\dots\}$ being the sublattice of index $k$,
whereas $\J=\{1,2,\dots,k-1\}$.

We have therefore
\beq
I(z)=\frac{z^k}{1-z^k},\qquad J(z)=\frac{z-z^k}{1-z}.
\eeq

The formulas~(\ref{nres}) and~(\ref{zres}) respectively read
\beq
\N(z)=\frac{1-z^k}{1-z-z^k+z^{2k}},
\label{nk}
\eeq
\beq
Z(z,x)=\frac{1-z^k}{1-z-x^kz^k+x^kz^{2k}}.
\label{zk}
\eeq
The first of these results was derived by a purely combinatorial approach in~\cite{dos2}.
The numerators and denominators of~(\ref{nk}) and~(\ref{zk}) have the common root $z=1$.
As a consequence, the degree of the denominators
\beqa
D(z)=1-z-z^k+z^{2k},
\nonumber\\
D(z,x)=1-z-x^kz^k+x^kz^{2k}
\label{dk}
\eeqa
can be reduced to $\Delta=2k-1$.

The relationship between the parameter $x$
and the smallest root $z_0(x)$ of the polynomial $D(z,x)$ takes the form
\beq
x^k=\frac{1-z_0(x)}{z_0(x)^k(1-z_0(x)^k)}.
\label{xk}
\eeq
The expression~(\ref{rhores}) for the density $\rho$ therefore reads
\beq
\rho=\frac{k(1-z_0(x))(1-z_0(x)^k)}{k+(1-k)z_0(x)-2kz_0(x)^k+(2k-1)z_0(x)^{k+1}}.
\label{rhok}
\eeq
The expression~(\ref{sres}) for the configurational entropy simplifies to
\beq
S(\rho)=-(1-\rho)\ln z_0(x)+\frac{\rho}{k}\ln\frac{1-z_0(x)^k}{1-z_0(x)},
\label{sk}
\eeq
where $\rho$ varies between the extremal values
\beq
\rho_\min=\frac{k}{2k-1},\qquad\rho_\max=1.
\eeq

The maximum of the configurational entropy reads $S_\star=-\ln z_\star$ (see~(\ref{szstar})),
where $z_\star=z_0(1)$ obeys $D(z_\star)=0$ (see~(\ref{dk})), i.e.,
\beq
1-z_\star-z_\star^k+z_\star^{2k}=0.
\eeq
The corresponding particle density reads
\beq
\rho_\star=c_1=\frac{kz_\star^k(1-z_\star^k)}{z_\star+kz_\star^k-2kz_\star^{2k}}.
\eeq
This expression can be obtained either from~(\ref{xk}) and~(\ref{rhok}),
or by using the definition~(\ref{fdef}) and the expansion~(\ref{fexpan})
of the configurational free energy $F(\beta)$.
The second approach yields the higher cumulant amplitudes $c_n$ as well.
We mention here the expression of the variance amplitude for further reference:
\beq
c_2=\frac{k^2z_\star^k(1-z_\star^k)(z_\star^2-z_\star^{k+1}+z_\star^{2k+1}-k^2z_\star^{3k})}
{(z_\star+kz_\star^k-2kz_\star^{2k})^3}.
\label{kc2}
\eeq

The above expressions somehow simplify in the case of dimers ($k=2$)
and in the regime where $k$ is very large.
These two situations are studied in detail below.

\subsection{The case of dimers $(k=2)$}

The general theory simplifies as follows in the case of dimers.
First,~(\ref{nres}) reads
\beq
\N(z)=\frac{1+z}{1-z^2-z^3},
\eeq
implying that the total numbers $\N_N$ of configurations obey the recursion
\beq
\N_N=\N_{N-2}+\N_{N-3},
\label{npado}
\eeq
defining the Padovan numbers $P_N$.
These integers are listed in entry A000931 of the OEIS~\cite{OEIS},
together with many formulas and references.
The initial values $\N_0=\N_1=\N_2=1$ exactly lead to
\beq
\N_N=P_N.
\label{npres}
\eeq
The recursion~(\ref{npado}) and its solution~(\ref{npres})
were obtained by combinatorial approaches in~\cite{dos1,kld}.

The case of dimers is also
the only situation where~(\ref{xk}) and~(\ref{rhok}) yield rational expressions
for $x^2$ and $z_0(x)$ in terms of the density $\rho$:
\beq
x^2=\frac{(3\rho-2)^3}{4\rho(1-\rho)^2},\qquad
z_0(x)=\frac{2(1-\rho)}{3\rho-2}.
\label{ratpar}
\eeq
As a consequence,~(\ref{sk}) translates to
\beq
S(\rho)=-\frac{1}{2}(3\rho-2)\ln(3\rho-2)-(1-\rho)\ln(2(1-\rho))+\frac{1}{2}\,\rho\ln\rho,
\label{sk2}
\eeq
where the density $\rho$ varies between $\rho_\min=2/3$ and $\rho_\max=1$.

The quantity $z_\star=z_0(1)$ obeys the cubic equation
\beq
z_\star^3+z_\star^2-1=0,
\label{zdefd}
\eeq
and so
\beq
z_\star
=\frac{1}{6}\bigl((100+12\sqrt{69})^{1/3}+(100-12\sqrt{69})^{1/3}-2\bigr)
\approx0.754877
\eeq
is the reciprocal of the so-called plastic number
\beq
\phi
=\frac{1}{6}\bigl((108+12\sqrt{69})^{1/3}+(108-12\sqrt{69})^{1/3}\bigr)
\approx1.324717,
\eeq
obeying
\beq
\phi^3-\phi-1=0,
\eeq
and describing the asymptotic growth of Padovan numbers: $P_N\sim\phi^N$.
We have
\beqa
S_\star
&=&-\ln z^\star=\ln\phi\approx0.281199,
\nonumber\\
\rho_\star
&=&\frac{2(z^\star+1)}{3z^\star+2}=\frac{2(3z_\star^2+z_\star+7)}{23}\approx0.822991,
\nonumber\\
c_2
&=&\frac{4(29z_\star^2+2z_\star-9)}{529}\approx0.068318.
\eeqa
The first expression of $\rho_\star$ has been derived from~(\ref{ratpar}),
whereas the second one has been obtained by reducing the first one
by means of the definition~(\ref{zdefd}) of~$z_\star$.
The expression of $c_2$ has been derived from~(\ref{kc2}) by the same procedure.
Any rational expression in $z_\star$ can indeed be reduced to a quadratic polynomial.

\subsection{The regime of large $k$}

The theory also simplifies for large $k$.
There, a scaling regime where~$z_0(x)$ is close to unity
dictates both the behavior of the cumulant amplitudes $c_n$
and the form of the configurational entropy $S(\rho)$ in the vicinity of its maximum~$S_\star$.
In order to explore this regime, we set
\beq
z_0(x)=\exp\left(-\frac{u(x)}{k}\right).
\eeq
Terms of relative order $1/k$ will be consistently neglected throughout the following.
The key equations~(\ref{xk}),~(\ref{rhok}) and~(\ref{sk}) respectively simplify to
\beq
x^k\approx\frac{1}{k}\,u(x)\e^{u(x)},
\label{xsca}
\eeq
\beq
\rho\approx\frac{u(x)}{u(x)+1},
\label{rsca}
\eeq
\beq
S(\rho)\approx\frac{1}{k}\,\frac{u(x)}{u(x)+1}\left(1+\ln\frac{k}{u(x)}\right).
\label{ssca}
\eeq
The last two relations can be combined to give
\beq
S(\rho)\approx\frac{\rho}{k}\,\left(1+\ln\frac{k(1-\rho)}{\rho}\right).
\label{sscarho}
\eeq

Let us consider first the maximum $S_\star$ of the configurational entropy
and the corresponding value $\rho_\star$ of the particle density.
The most direct approach consists in setting $x=1$ in~(\ref{xsca}).
We thus find that $u_\star=u(1)$ obeys
\beq
u_\star\e^{u_\star}\approx k,
\label{ustar}
\eeq
whereas~(\ref{rsca}) and~(\ref{ssca}) respectively become
\beq
\rho_\star\approx\frac{u_\star}{u_\star+1},
\label{rhostar}
\eeq
\beq
S_\star\approx\frac{u_\star}{k}.
\label{sstar}
\eeq
The solution to~(\ref{ustar}) reads
\beq
u_\star\approx W(k),
\label{ustarw}
\eeq
in terms of the Lambert $W$ function.
To leading order as $k$ is very large, we have
\beq
u_\star\approx\ln k,
\label{ulead}
\eeq
and therefore
\beq
\rho_\star\approx1-\frac{1}{\ln k},
\label{rlead}
\eeq
\beq
S_\star\approx\frac{\ln k}{k}.
\label{slead}
\eeq
The above estimates involving logarithmic factor will be commented
and illustrated at the end of this section.
For the time being,
let us show that they can be turned into full asymptotic expansions
in inverse powers of $\ln k$.
Introducing the notations
\beq
\lam=\ln k,\qquad
\mu=\ln\lam=\ln\ln k,
\eeq
and setting
\beq
u_\star=\lam-\mu+\eps,
\label{ustars}
\eeq
the expression~(\ref{ustar}) translates to
\beq
\eps\approx-\ln\left(1+\frac{\eps-\mu}{\lam}\right).
\eeq
This equation can be solved by iteration, yielding the expansion
\beq
\eps\approx\frac{\mu}{\lam}+\frac{\mu(\mu-2)}{2\lam^2}
+\frac{\mu(2\mu^2-9\mu+6)}{6\lam^3}+\cdots
\label{epss}
\eeq
Inserting~(\ref{ustars}),~(\ref{epss}) into~(\ref{rhostar}),~(\ref{sstar}),
we obtain the expansions
\beqa
\rho_\star\approx1-\frac{1}{\lam}-\frac{\mu-1}{\lam^2}-\frac{\mu^2-3\mu+1}{\lam^3}
-\frac{2\mu^3-11\mu^2+12\mu-2}{2\lam^4}+\cdots,
\nonumber\\
\label{rhoser}
\eeqa
\beq
S_\star\approx\frac{1}{k}\left(\lam-\mu+\frac{\mu}{\lam}+\frac{\mu(\mu-2)}{2\lam^2}
+\frac{\mu(2\mu^2-9\mu+6)}{6\lam^3}+\cdots\right).
\label{sser}
\eeq
All terms of the above asymptotic expansions can be trusted.
The present approach indeed neglects corrections of relative order $1/k$,
which are exponentially small in $\lam$.
Conversely,~(\ref{ustar}),~(\ref{rhostar}) and~(\ref{sstar})
can be thought of as all-order resummations of the above asymptotic expansions.

The shape of the entropy~$S(\rho)$ in the vicinity of its maximum $S_\star$
can be investigated by simplifying the expression~(\ref{sscarho})
in the regime where $1-\rho$ is proportional to $1/\lam$.
This scaling is suggested by the estimate~(\ref{rlead}) of $\rho_\star$.
Introducing the variable
\beq
\xi=\lam(1-\rho),
\eeq
we recover from~(\ref{sscarho}) the first two terms of the expansion~(\ref{sser}) of $S_\star$,
as well as a scaling expression for the large-deviation function
$\Sigma(\rho)=S_*-S(\rho)$ to leading order in $\lam$, namely
\beq
\Sigma(\rho)\approx\frac{\Phi(\xi)}{k},
\label{sigphi}
\eeq
where the scaling function reads
\beq
\Phi(\xi)=\xi-1-\ln\xi.
\label{phi}
\eeq
This function vanishes for $\xi_*=\lam(1-\rho_\star)=1$,
in agreement with~(\ref{rlead}) and with the first two terms of~(\ref{rhoser}),
around which it behaves quadratically as
\beq
\Phi(\xi)\approx\frac{(\xi-1)^2}{2},
\eeq
so that
\beq
\Sigma(\rho)\approx\frac{\lam^2(\rho-\rho_\star)^2}{2k},
\eeq
in agreement with~(\ref{squadra}), and with~(\ref{c2sca}) to leading order in $\lam$.

The behavior of the cumulant amplitudes $c_n$ can also be investigated along the same lines.
Setting
\beq
\beta=\frac{\sigma}{k},\qquad\hbox{i.e.,}
\quad x=\exp\left(\frac{\sigma}{k}\right),
\label{xsig}
\eeq
where the parameter $\sigma$ is independent of $k$,~(\ref{xsca}) becomes
\beq
u(x)\e^{u(x)}\approx k\e^\sigma,
\label{usig}
\eeq
hence
\beq
u(x)\approx W(k\e^\sigma).
\eeq
The configurational free energy introduced in~(\ref{fdef}) reads
\beq
F(\beta)\approx\frac{u(x)-u_\star}{k}.
\eeq
Expanding $u(x)$ in powers of $\sigma$
and identifying coefficients with~(\ref{fexpan}), we obtain
\beq
c_1=\rho_\star\approx\frac{u_\star}{u_\star+1}
\label{c1sca}
\eeq
(see~(\ref{rhostar}),~(\ref{rlead}),~(\ref{rhoser})), as well as
\beqa
c_2
&\approx&\frac{k u_\star}{(u_\star+1)^3}
\nonumber\\
&\approx&\frac{k}{\lam^2}\left(1+\frac{2\mu-3}{\lam}+\frac{3\mu^2-11\mu+6}{\lam^2}+\cdots\right),
\label{c2sca}
\\
c_3
&\approx&-\frac{k^2 u_\star(2u_\star-1)}{(u_\star+1)^5}
\nonumber\\
&\approx&-\frac{2k^2}{\lam^3}\left(1+\frac{6\mu-11}{2\lam}
+\frac{12\mu^2-50\mu+35}{2\lam^2}+\cdots\right),
\eeqa
and so on.
To leading order in $\lam$, all cumulant amplitudes scale as
\beq
c_n-\delta_{n1}\approx(-1)^n(n-1)!\frac{k^{n-1}}{\lam^n}.
\label{clea}
\eeq
Inserting the above leading-order estimate
into~(\ref{fexpan}) and summing the series,
we obtain a scaling formula for the configurational free energy, namely
\beq
F(\beta)\approx\beta-\frac{1}{k}\,\ln\frac{\lam+k\beta}{\lam}.
\eeq
The expression~(\ref{rhof}) then yields
\beq
\rho\approx1-\frac{1}{\lam+k\beta}.
\eeq
Using~(\ref{sigap}),~(\ref{sigr}),
we recover the scaling expression~(\ref{sigphi}),~(\ref{phi})
for the large-deviation function $\Sigma(\rho)$.

Let us now comment and illustrate the above results.
The most striking feature of the large-$k$ regime
is the occurrence of logarithms
in the leading-order estimates~(\ref{rlead}) and~(\ref{slead}).
These logarithmic corrections originate in
the following peculiar feature of the large-$k$ regime.
The length of any cluster of empty sites between two sequences of $k$-mers
may take a very large number $(k-1)$ of different values.
This proliferation of the possible numbers of consecutive empty sites
affects the renewal formalism as follows.
In the naive scaling regime where $x^k$ and $z^k$ are of order unity,
$I(xz)$ remains of order unity, whereas~$J(z)$ grows proportionally to $k$.
This fundamental dissymmetry between occupied and empty sites
manifests itself by the presence of a factor $k$ in the right-hand side of~(\ref{ustar}),
whose solution scales as~(\ref{ulead}).
It is therefore at the origin of the observed logarithmic corrections to scaling.
The slow convergence of the most probable density~$\rho_\star$ to
the largest possible density $\rho_\max=1$
is the foremost consequence of this mechanism.

\begin{figure}
\begin{center}
\includegraphics[angle=0,width=0.6\linewidth,clip=true]{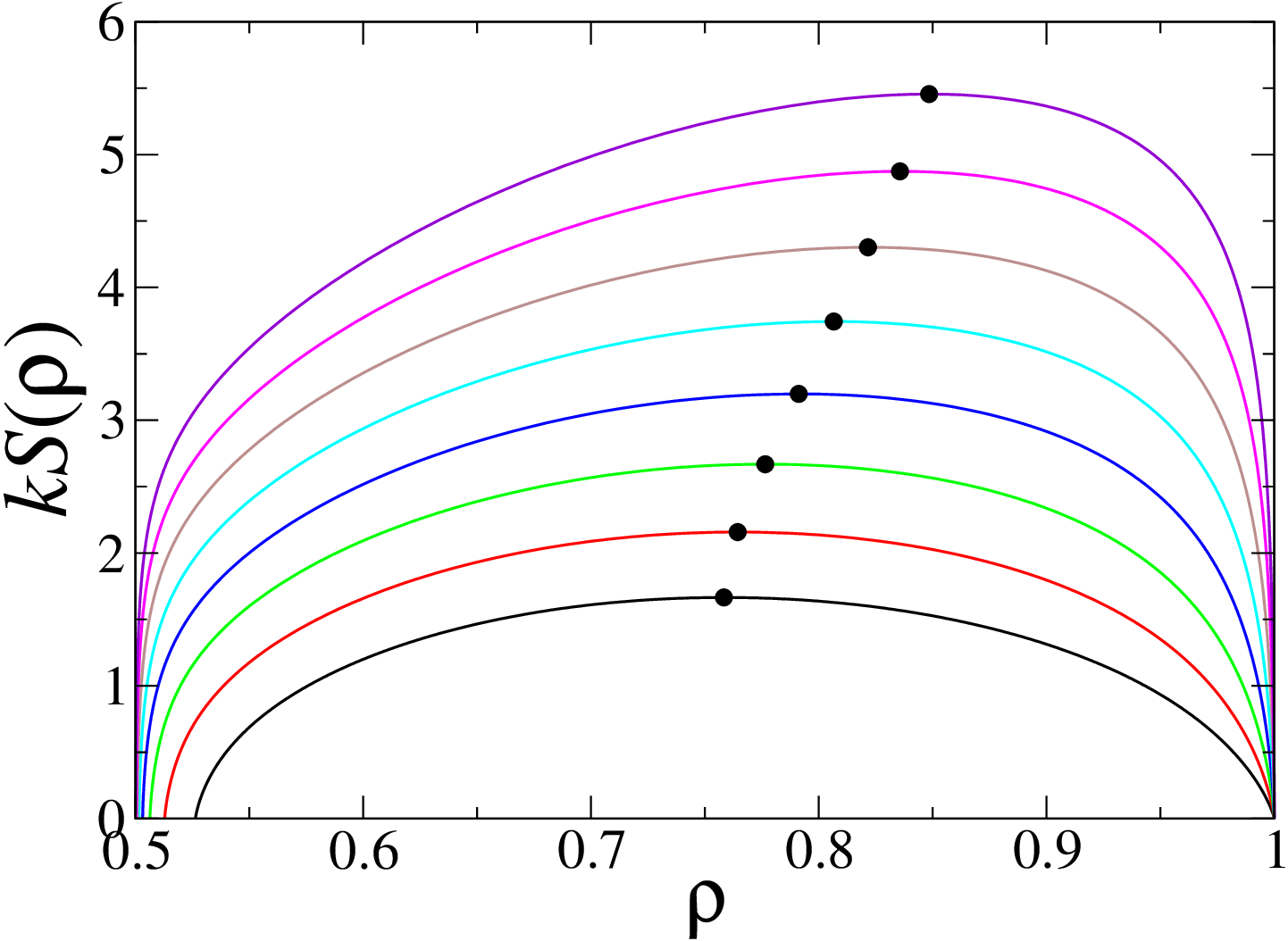}
\caption{\small
Configurational entropy $S(\rho)$ of the $k$-mer problem,
multiplied by $k$ and plotted against particle density $\rho$.
Bottom to top: $k=10$, 20, 40, 80, 160, 320, 640 and 1280.
Symbols: coordinates $(\rho_\star, kS_\star)$ of the maxima
of the entropy curves.}
\label{kent}
\end{center}
\end{figure}

Figure~\ref{kent} shows plots of the configurational entropy $S(\rho)$, multiplied by $k$,
against the particle density $\rho$ for a broad range of values of $k$,
equally spaced on a logarithmic scale.
The heights of the maxima $kS_\star$ (black symbols) are roughly equidistant,
in agreement with the prediction~(\ref{slead}).
For modest values of~$k$, the entropy curves are roughly symmetric.
Accordingly, the most probable densities $\rho_\star$ are not far from the middle
of the interval of values of $\rho$, i.e., $\rho_{\rm mid}=3/4$.
The asymptotic prediction~(\ref{rlead}) for the most probable density $\rho_\star$
implies that the entropy curves eventually become very asymmetric for very large values of $k$,
with their maxima approaching the upper edge of the interval at logarithmic speed.
This asymmetry is indeed observed to set in very slowly.

\begin{figure}
\begin{center}
\includegraphics[angle=0,width=0.6\linewidth,clip=true]{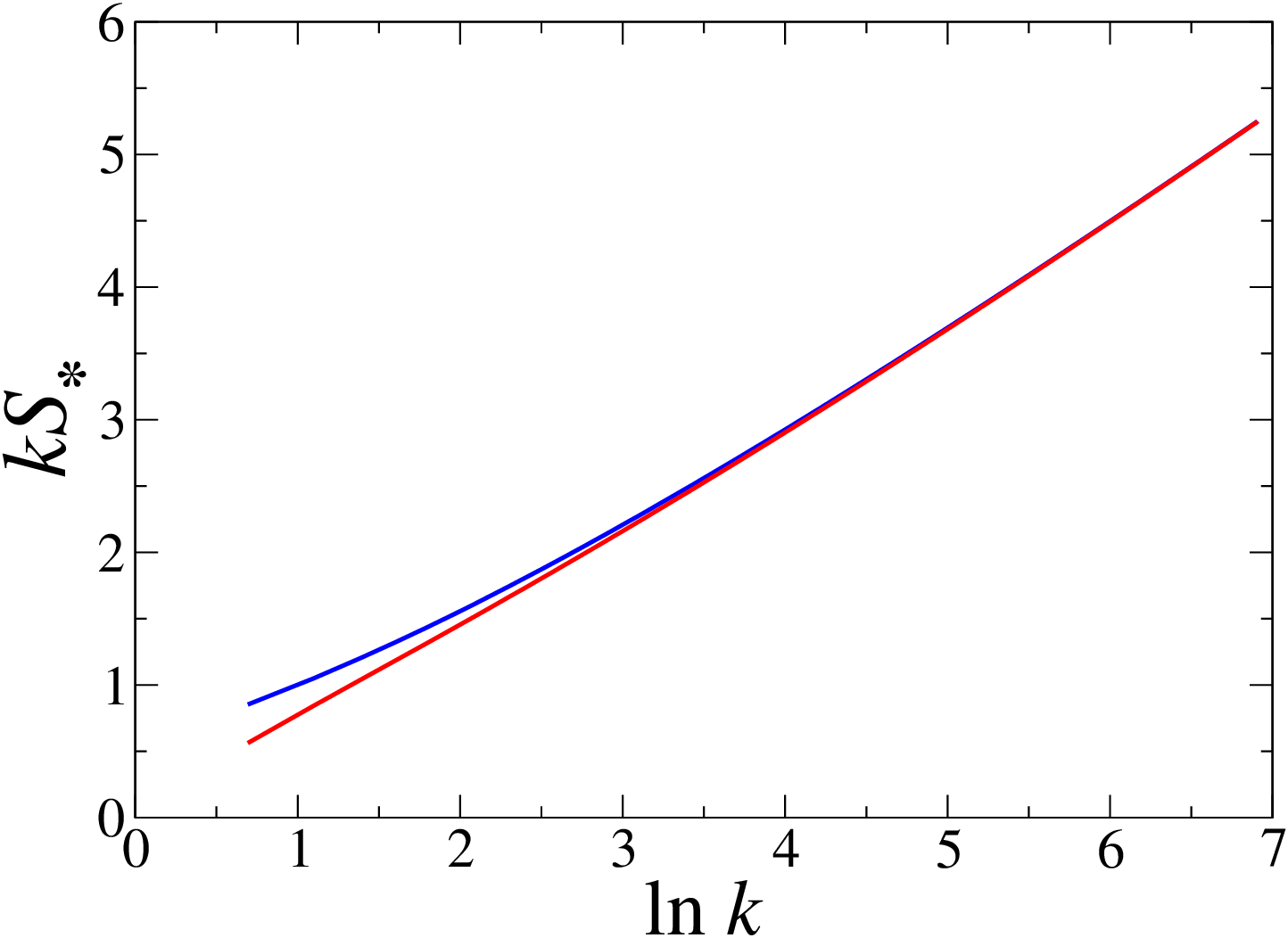}
\caption{\small
Maximal configurational entropy $S_\star$ of statistical ensemble
of blocked $k$-mer configurations,
multiplied by $k$ and plotted against $\ln k$.
Red curve: values of $kS_\star$ for finite $k$.
Blue curve: large-$k$ estimate (see~(\ref{sstar}),~(\ref{ustarw})).}
\label{ksstar}
\end{center}
\end{figure}

Figure~\ref{ksstar} shows the maximal entropy $S_\star$,
multiplied by $k$ and plotted against $\ln k$.
The values of $kS_\star$ (red)
are compared to the all-order large-$k$ estimate $u_\star$
(see~(\ref{sstar}), (\ref{ustarw})) (blue).
The latter estimate converges very fast to the data,
in agreement with the expectation that this convergence is in $1/k$,
roughly speaking, i.e., exponentially fast at the scale of the plot.

\begin{figure}
\begin{center}
\includegraphics[angle=0,width=0.6\linewidth,clip=true]{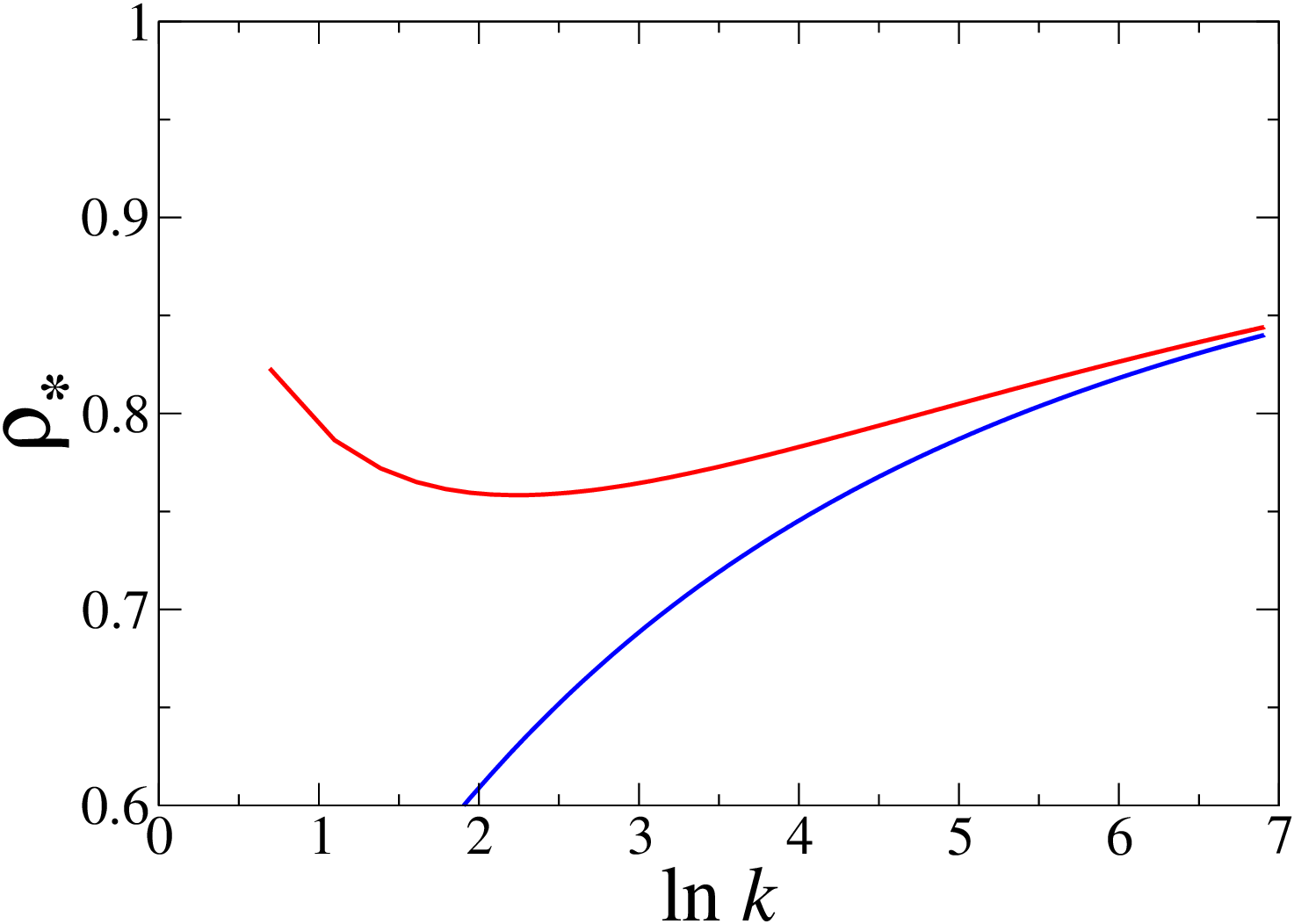}
\caption{\small
Most probable density $\rho_\star$ of statistical ensemble
of blocked $k$-mer configurations, against $\ln k$.
Red curve: values of $\rho_\star$ for finite $k$.
Blue curve: large-$k$ estimate (see~(\ref{rhostar}),~(\ref{ustarw})).}
\label{krstar}
\end{center}
\end{figure}

\begin{figure}
\begin{center}
\includegraphics[angle=0,width=0.6\linewidth,clip=true]{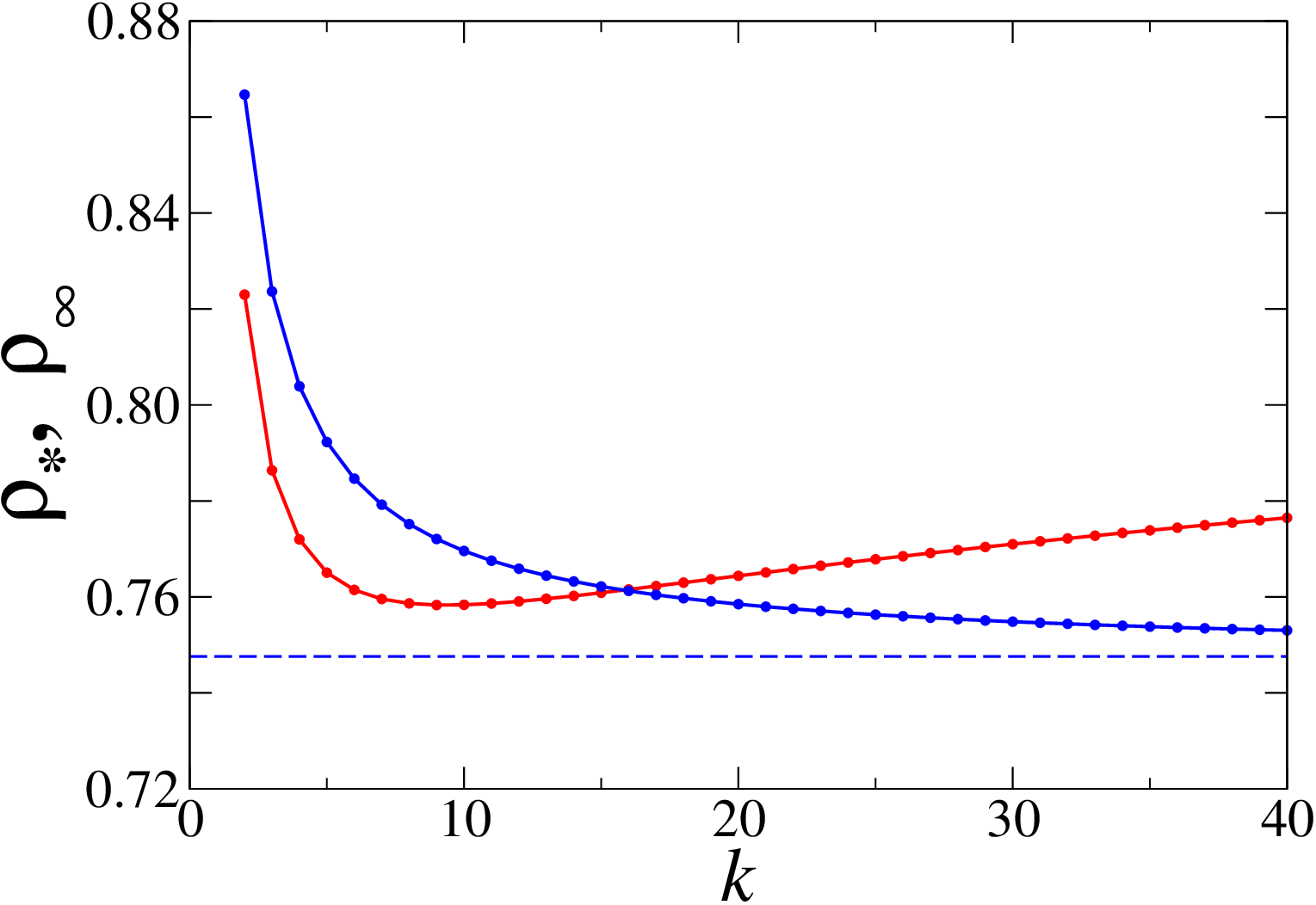}
\caption{\small
Particle densities of blocked $k$-mer configurations, against~$k$.
Red curve: most probable density $\rho_\star$ of the statistical ensemble.
Blue curve: dynamical density $\rho_\infty$ (see~(\ref{rhody})).
Horizontal dashed line: R\'enyi parking constant $R$ (see~(\ref{ren})).}
\label{krsd}
\end{center}
\end{figure}

Figure~\ref{krstar} shows the most probable density $\rho_\star$ plotted against $\ln k$.
This density (red) exhibits a non-monotonic dependence on $k$.
More details will be given in the discussion of figure~\ref{krsd}.
The data are compared to the all-order large-$k$ estimate
(see~(\ref{rhostar}),~(\ref{ustarw})) (blue).
The accuracy of the latter estimate is far less impressive than for $S_\star$
(see figure~\ref{ksstar}).

Figure~\ref{krsd} presents a comparison
between the a priori (or static) density of the problem of $k$-mers,
i.e., the most probable density $\rho_\star$ of the statistical ensemble
of blocked configurations investigated above (red),
and the final (or dynamical) density~$\rho_\infty$ of the blocked configurations
reached by the deposition dynamics starting from an empty lattice (blue).

The dynamical density is exactly known for all values of $k$~\cite{ghh,beg,bonnier,kld}:
\beq
\rho_\infty=k\int_0^1\exp\Biggl(-2\sum_{j=1}^{k-1}\frac{1-y^j}{j}\Biggr)\,\dd y.
\label{rhody}
\eeq
For $k=2$, we recover the result by Flory for the dimer deposition problem~\cite{flory}:
\beq
\rho_\infty=1-\e^{-2}\approx0.864664.
\eeq
The modest range $(k\le40)$ used in figure~\ref{krsd}
allows a good visibility of the following traits.
The dynamical density $\rho_\infty$ is a monotonically decreasing function of~$k$,
whereas the static density $\rho_\star$ exhibits a non-monotonic dependence on $k$,
reaching its minimum $\rho_\star\approx0.758316$ for $k=9$,
as already noticed in~\cite{dos2}.
Both particle densities cross between 15 and~16.
As a consequence, for $k\le 15$ we have $\rho_\infty>\rho_\star$,
as in most common RSA and related models~\cite{epjst},
whereas the reverse inequality $\rho_\infty<\rho_\star$ holds for $k\ge 16$.
For large~$k$, the dynamical density converges
to the celebrated R\'enyi parking constant~\cite{renyi}:
\beq
R=\int_0^\infty\exp\left(-2\int_0^y\frac{1-\e^{-x}}{x}\,\dd x\right)\dd y\approx0.747597,
\label{ren}
\eeq
describing the final density of the blocked configurations
for the RSA of unit intervals on the continuous line.
The corrections to the above limit have been shown~\cite{beg}
to be given by a power series in inverse powers of $k$:
\beq
\rho_\infty=R+\frac{R_1}{k}+\frac{R_2}{k^2}+\cdots,
\eeq
with $R_1\approx0.216181$, $R_2\approx0.036255$.
The convergence of $\rho_\infty$ to its limit~$R$ (horizontal dashed line)
is much faster than the logarithmically slow convergence of $\rho_\star$
to the limit $\rho_\max=1$.

\section{Rydberg atoms}
\label{rydberg}

This section is devoted to configurations of assemblies of trapped ultracold Rydberg atoms.
In the simple one-dimensional setting described in the introduction,
Rydberg atoms are viewed as particles occupying the sites of a one-dimensional optical lattice,
with a constraint stemming from the Rydberg blockade,
namely that each occupied site must have at least $b$ empty sites on either side.
The integer $b\ge1$ is referred to as the blockade range of the model.
Our aim is to investigate the ensemble of all blocked configurations,
where no single atom can be inserted any more~\cite{kld,crew}.

A blocked configuration consists of isolated occupied sites (the Rydberg atoms)
separated by clusters of empty sites whose length is at least $b$,
in order to obey the blockade constraint, and at most~$2b$,
since an extra Rydberg atom could be inserted in the middle of an empty range of size $2b+1$.
On a finite sample with open boundary conditions,
there can be at most~$b$ empty sites on the left of the first occupied one
and at most $b$ empty sites on the right of the last occupied one.
Blocked configurations of Rydberg atoms can therefore
be described in terms of independent clusters.
Within the renewal approach of section~\ref{renewal},
clusters of occupied and empty sites respectively correspond to
$\I=\{1\}$ and $\J=\{b,b+1,\dots,2b\}$.
The renewal approach must however be complemented
in order to take boundary conditions into account,
namely the presence of empty sites near both endpoints,
whose numbers belong to the set $\B=\{0,1,\dots,b\}$.
The associated generating series read
\beq
I(z)=z,\qquad J(z)=\frac{z^b(1-z^{b+1})}{1-z},\qquad B(z)=\frac{1-z^{b+1}}{1-z}.
\label{rsers}
\eeq
The formula~(\ref{nres}) becomes
\beqa
\N(z)&=&1+B(z)I(z)B(z)+B(z)I(z)J(z)I(z)B(z)+\cdots
\nonumber\\
&=&\frac{1+I(z)(B(z)^2-J(z))}{1-I(z)J(z)}.
\label{nfor}
\eeqa
Similarly, the formula~(\ref{zres}) becomes
\beq
Z(z,x)=\frac{1+I(xz)(B(z)^2-J(z))}{1-I(xz)J(z)}.
\label{zfor}
\eeq
Using~(\ref{rsers}), the expressions~(\ref{nfor}) and~(\ref{zfor}) read explicitly
\beqa
\N(z)=\frac{1-z+z^2-z^{b+1}-z^{b+2}+z^{2b+2}}{(1-z)(1-z-z^{b+1}+z^{2b+2})},
\label{nry}
\\
Z(z,x)=\frac{1+(x-2)z+z^2-xz^{b+1}-xz^{b+2}+xz^{2b+2}}{(1-z)(1-z-xz^{b+1}+xz^{2b+2})}.
\label{zry}
\eeqa
These rational expressions are not singular at $z=1$,
so that their denominators actually read
\beqa
D(z)=1-z-z^{b+1}+z^{2b+2},
\nonumber\\
D(z,x)=1-z-xz^{b+1}+xz^{2b+2}.
\label{dr}
\eeqa
These denominators do not depend on the specific boundary conditions.
This is to be expected, as they encode the properties of the system
in the thermodynamic limit.
They can be mapped onto the denominators of the $k$-mer problem
(see~(\ref{dk})) by setting
\beq
k=b+1,
\label{kb}
\eeq
and changing $x$ into $x^k$.

Figure~\ref{seqs} (see section~\ref{kmer})
demonstrates the equivalence underlying the above observation
at the level of single configurations.
A Rydberg atom followed by $b$ empty sites (to its right) can be mapped onto a $k$-mer,
where $b$ and $k$ are related by~(\ref{kb}).
The numbers $M_\ra$ of Rydberg atoms and~$M$ of particles in the $k$-mer problem,
and the corresponding densities $\rho_\ra$ and $\rho$,
are therefore related by
\beq
M_\ra=\frac{M}{k},\qquad
\rho_\ra=\frac{\rho}{k}.
\eeq
The mapping between $k$-mers and Rydberg atoms with blockade range $b$ is however not unique.
For instance, a $k$-mer can equally well represent a Rydberg atom preceded by $b$ empty sites
(to its left).
As a consequence of this non-uniqueness, the equivalence between both models
does not exactly hold on finite systems with prescribed boundary conditions.
This explains why
the denominators~(\ref{dr}) and~(\ref{dk}) are simply related to each other,
as observed above,
whereas the full generating series~(\ref{nry}),~(\ref{zry})
are not simply related to~(\ref{nk}),~(\ref{zk}).

All results derived in section~\ref{kmer}
concerning the $k$-mer problem in the thermo\-dynamic limit
apply {\it mutatis mutandis}
to blocked configurations of Rydberg atoms
on the formally infinite chain.
Hereafter we focus our attention
onto two quantities which are more specific to the latter problem.

\begin{figure}
\begin{center}
\includegraphics[angle=0,width=0.6\linewidth,clip=true]{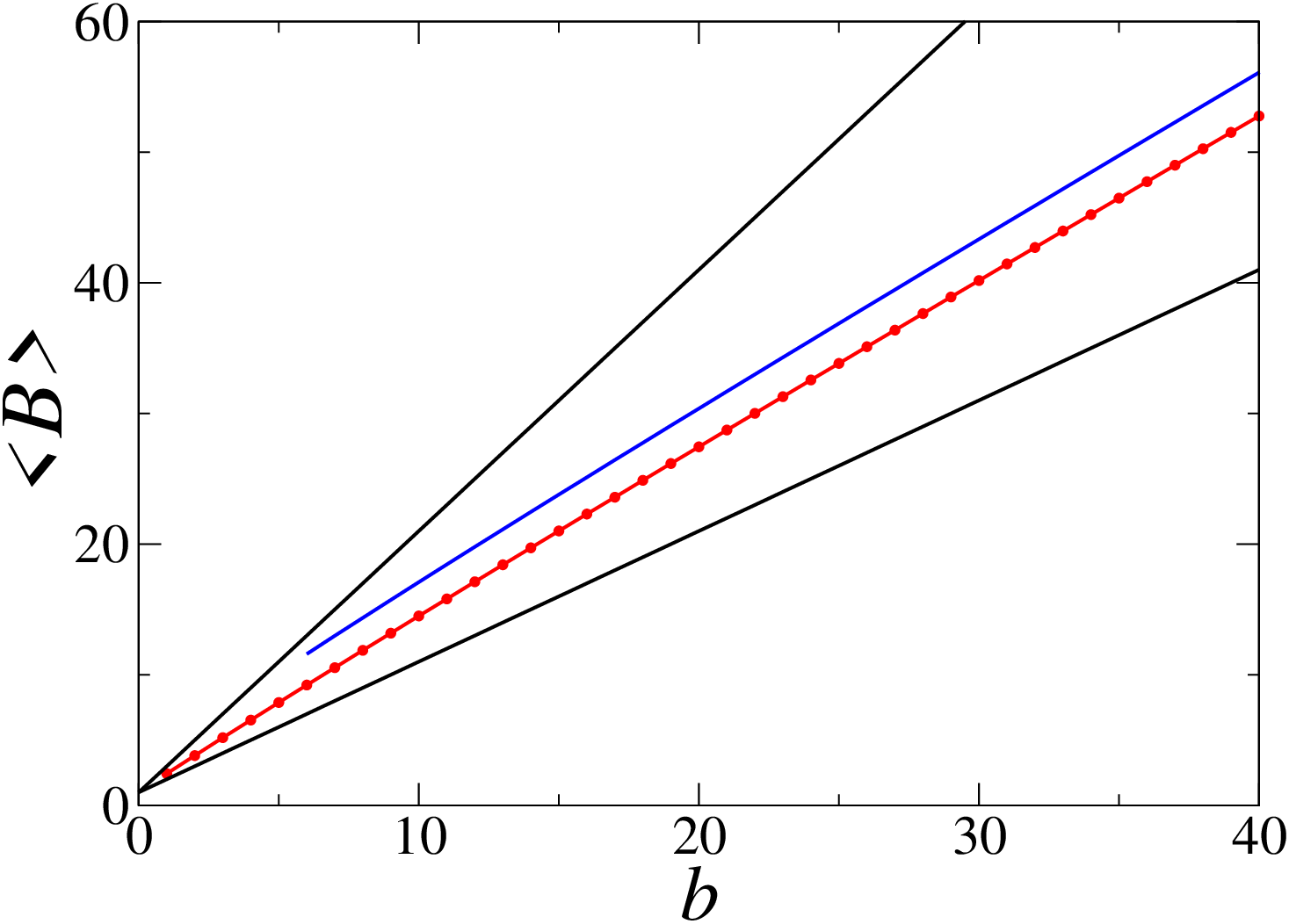}
\caption{\small
Mean interatomic distance $\mean{B}$ (see~(\ref{bdef}))
of blocked con\-figurations of Rydberg atoms,
against blockade range $b$.
Red line with symbols: values of $\mean{B}$.
Blue line: large-$k$ estimate (see~(\ref{rhostar}), (\ref{ustarw})).
Black lines: extremal values $B_\min=b+1$ and $B_\max=2b+1$.}
\label{bra}
\end{center}
\end{figure}

The first observable of interest is the mean distance $\mean{B}$
between successive Rydberg atoms along the chain, measured in lattice spacings.
In the statistical ensemble of blocked configurations,
this mean distance reads
\beq
\mean{B}=\frac{1}{(\rho_\ra)_\star}=\frac{k}{\rho_\star},
\label{bdef}
\eeq
where quantities entering the rightmost side pertain to the $k$-mer problem,
studied in section~\ref{kmer}.
The mean interatomic distance $\mean{B}$ only depends on the blockade range~$b$.
It is given in table~\ref{bq} up to $b=10$,
and plotted against $b$ in figure~\ref{bra} up to $b=40$.
Black straight lines show the extremal values $B_\min=b+1$ and $B_\max=2b+1$.
Actual values of $\mean{B}$ (red line with symbols)
are compared to the large-$k$ estimate for the ratio~$k/\rho_\star$
(see~(\ref{rhostar}),~(\ref{ustarw})) (blue line).
The latter estimate has the asymptotic expansion (see~(\ref{rhoser}))
\beq
\mean{B}=b\left(1+\frac{1}{\lam}+\frac{\mu}{\lam^2}+\frac{\mu(\mu-1)}{\lam^3}+\cdots\right),
\eeq
with
\beq
\lam=\ln b,\qquad
\mu=\ln\lam=\ln\ln b,
\eeq
implying that $\mean{B}\approx b(1+1/\ln b)$ is hardly larger than $B_\min\approx b$ at large $b$.
The growth of $\mean{B}$ is already rather well represented by the above estimate
for the modest range of $b$ shown in figure~\ref{bra}.

\begin{table}
\begin{center}
$$
\begin{array}{|c|c|c|}
\hline
b & \mean{B} & Q \\
\hline
1 & \;2.430159 & -\,0.958493 \\
2 & \;3.814962 & -\,0.955953 \\
3 & \;5.181490 & -\,0.955998 \\
4 & \;6.535473 & -\,0.956436 \\
5 & \;7.879669 & -\,0.956919 \\
6 & \;9.215803 & -\,0.957378 \\
7 & 10.545079 & -\,0.957798 \\
8 & 11.868394 & -\,0.958181 \\
9 & 13.186446 & -\,0.958531 \\
10 & 14.499793 & -\,0.958852 \\
\hline
\end{array}
$$
\caption{
Mean interatomic distance $\mean{B}$ (see~(\ref{bdef}))
and Mandel $Q$ parameter (see~(\ref{qasy})) in the statistical ensemble
of blocked configurations of Rydberg atoms with blockade range~$b$, up to $b=10$.}
\label{bq}
\end{center}
\end{table}

\begin{figure}
\begin{center}
\includegraphics[angle=0,width=0.6\linewidth,clip=true]{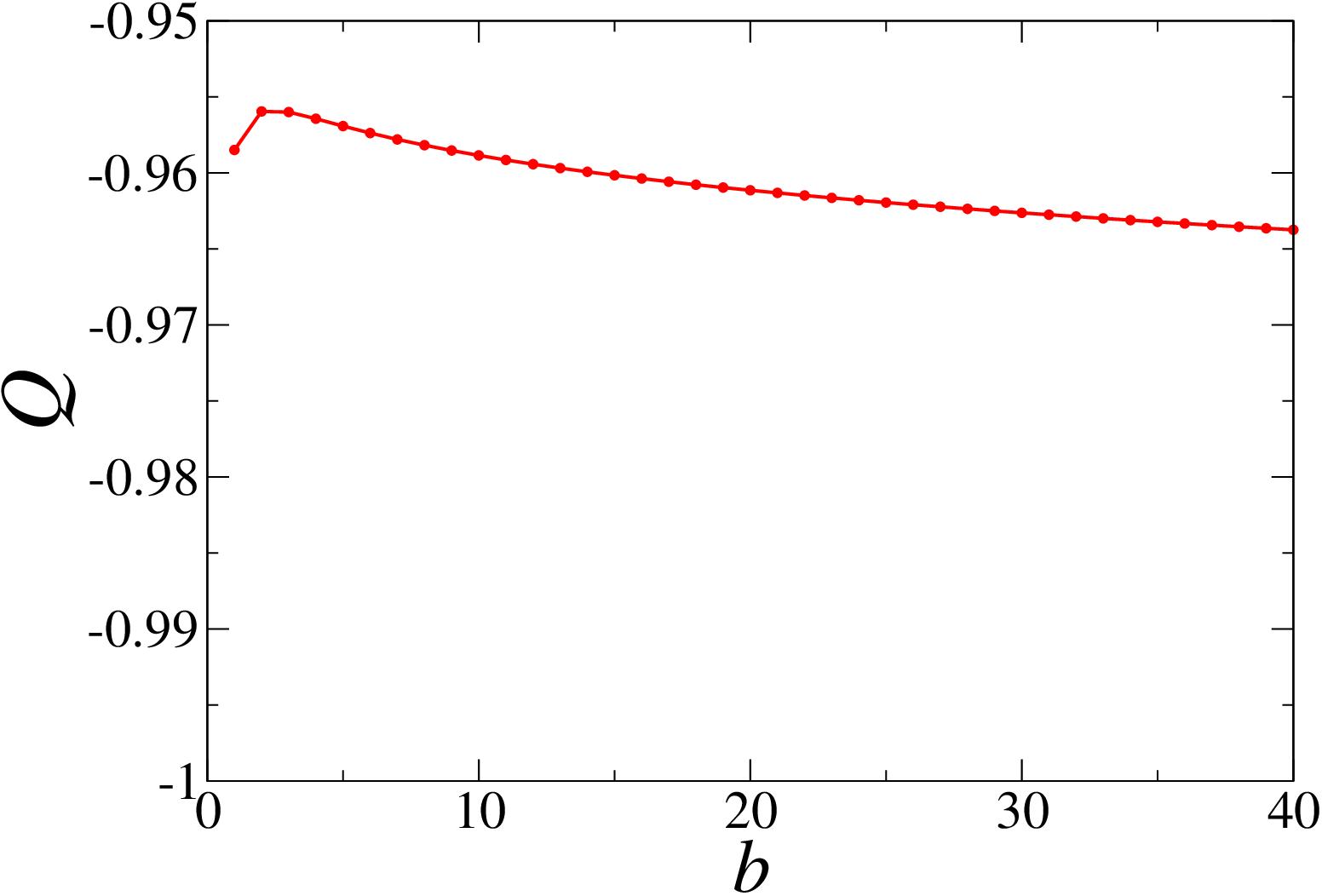}
\caption{\small
Mandel $Q$ parameter (see~(\ref{qasy})) of blocked configurations of Rydberg atoms,
against blockade range $b$.}
\label{qra}
\end{center}
\end{figure}

The second observable of interest is the so-called Mandel $Q$ parameter~\cite{mandel}:
\beq
Q=\frac{\mean{M_\ra^2}_c}{\mean{M_\ra}}-1.
\eeq
This parameter is commonly used in quantum optics and atomic physics.
It provides a measure of the deviation with respect to a Poissonian distribution,
for which $Q=0$.
In the present case, the Mandel $Q$ parameter
has a well-defined value in the thermodynamic limit,
which only depends on $b$, namely (see~(\ref{c1c2}))
\beq
Q=\frac{c_2}{kc_1}-1.
\label{qasy}
\eeq
Quantities in the right-hand side again pertain to the $k$-mer problem,
studied in section~\ref{kmer}.
The above $Q$ parameter is given in table~\ref{bq} up to $b=10$,
and plotted against $b$ in figure~\ref{qra} up to $b=40$.
It turns out that $Q$ is always near $Q_\min=-1$,
and has a very weak non-monotonic dependence on $b$, with a maximum at $b=2$.
This observation is in qualitative agreement with various experiments on Rydberg atoms,
where rather large negative values of $Q$ have been reported,
testifying strongly sub-Poissonian statistics~\cite{swm,lbr,vhb,hgs}.
The asymptotic estimate of $Q$ at large $b$,
namely (see~(\ref{ustars}), (\ref{epss}), (\ref{c1sca}), (\ref{c2sca}))
\beq
Q\approx\frac{1}{(u_\star+1)^2}-1
=-\,1+\frac{1}{\lam^2}+\frac{2(\mu-1)}{\lam^3}+\cdots,
\eeq
exhibits a logarithmically slow convergence toward $Q_\min=-1$.
The above estimate is however too large (i.e., too far from $Q_\min$)
to be visible in figure~\ref{qra}.
It is altogether of little practical use, except for extremely large $b$.

\section{Discussion}
\label{disc}

In this paper we have put forward an alternative method to investigate
statistical ensembles of constrained configurations of particles
on the one-dimensional lattice.
The scope of this renewal approach is restricted to local constraints
which are expressible in terms of the lengths of clusters of occupied and empty sites.
Within this scope, the present renewal method is more systematic and easier to implement
than traditional approaches involving either direct combinatorial reasoning
or the transfer-matrix formalism.
The key formulas~(\ref{nres}) and~(\ref{zres}) are indeed explicit,
given the sets $\I$ and $\J$ of permitted cluster lengths.

In the broad class of rational models,
the complexity of a statistical ensemble is measured by
the degree $\Delta$ of the polynomial~$D(z)$.
In particular, the numbers of configurations on finite lattices
of length $N$ obey a linear recursion
with constant integer coefficients, whose number of terms is at most $\Delta+1$.
The integer $\Delta$ is the analogue of the dimension of the transfer matrix
in the transfer-matrix approach summarized in~\ref{tm}.
The renewal approach it however more straightforward,
as the transfer-matrix formalism requires both the choice of a relevant set
of partial partition functions and the explicit building of the corresponding transfer matrix.
Furthermore, the renewal approach extends to non-rational models,
which would require the construction of an infinite-dimensional transfer operator.

The renewal approach has been illustrated in detail on the $k$-mer deposition model
and on assemblies of trapped Rydberg atoms with blockade range $b$.
In the latter case, the presence of specific boundary conditions
led us to extend the renewal approach
and to generalize the formulas~(\ref{nres}) and~(\ref{zres})
to~(\ref{nfor}) and~(\ref{zfor}).
The statistical ensembles of blocked configurations of $k$-mers and of Rydberg atoms
are essentially equivalent to each other,
with the identification $k=b+1$.
Their most remarkable common feature is the occurrence of logarithmic corrections
in the regime where~$k$ or $b$ become large,
whose origin has been explained in detail.
In the $k$-mer model,
the most probable density $\rho_\star$ approaches its maximal value $\rho_\max=1$
very slowly, with a large correction scaling as $1/\ln k$.
In the case of Rydberg atoms,
the Mandel~$Q$ parameter is always very near its minimal value $Q_\min=-1$,
irrespective of the blockade range~$b$.
Its logarithmically slow asymptotic convergence to $Q_\min$
is therefore unobservable, except for irrealistically large $b$.

Finally, it would be desirable to extend
either the present renewal method or the transfer-matrix approach
to more complex geometries besides the one-dimensional case.
In this context, regular trees seem to be the most promising setting.

\ack
It is a pleasure for us to thank
Tomislav Do\v{s}li\'c, Mate Puljiz, Stjepan \v{S}ebek and Josip \v{Z}ubrini\'c
for fruitful exchanges during the concomitant elaboration of their preprint~\cite{crew}
and of the present work.
PLK is grateful to IPhT Paris-Saclay for excellent working conditions.

\section*{Data availability statement}

Data sharing not applicable to this article as no datasets were generated or analyzed
during the current study.

\appendix

\section{Several species of particles}

In this appendix we show that the renewal approach
can be generalized to models having several species of particles.
We denote these species by $a=1,\dots,A$.
Each site of the infinite half-line is occupied by a particle of some species.
If empty sites (holes) are permitted,
they are represented for definiteness by the last species $(a=A)$.

We consider the statistical ensemble of configurations
defined by the constraints that the lengths of clusters of $a$-particles
belong to some set $\I_a$, encoded in the generating series
\beq
I_a(z)=\sum_{i\in\I_a}z^i.
\eeq
The model considered in section~\ref{renewal} corresponds to $A=2$,
with $a=1$ and $a=2$ respectively corresponding to particles (occupied sites)
and holes (empty sites),
so that $\I_1=\I$, $\I_2=\J$.

The generating series of the numbers of configurations,
\beq
\N(z)=\sum_{N\ge0}\N_Nz^N,
\eeq
can be derived along the lines of~(\ref{nser})--(\ref{nres}).
We have
\beq
\N(z)=1+\sum_{a=1}^A\N_a(z),
\eeq
with
\beq
\N_a(z)=I_a(z)\Bigl(1+\sum_{b\ne a}\N_b(z)\Bigr)=I_a(z)(\N(z)-I_a(z)),
\eeq
hence
\beq
\N_a(z)=\frac{I_a(z)}{1+I_a(z)}\,\N(z),
\eeq
and finally
\beq
\N(z)=\frad{1}{1-\sum_{a=1}^A\frac{I_a(z)}{1+I_a(z)}}.
\label{nares}
\eeq
For $A=2$, we have
\beq
\N(z)=\frac{(1+I_1(z))(1+I_2(z))}{1-I_1(z)I_2(z)},
\eeq
in agreement with~(\ref{nres}).
For $A=3$, we have
\beqa
\N(z)=\frac{(1+I_1(z))(1+I_2(z))(1+I_3(z))}
{1-I_1(z)I_2(z)-I_1(z)I_3(z)-I_2(z)I_3(z)-2I_1(z)I_2(z)I_3(z)},
\nonumber\\
\eeqa
and so on.

The generating series of the numbers $\N_{N,M_1,\dots,M_A}$
of configurations with given prescribed numbers $M_a$
of $a$-particles can be derived along the same lines,
by attributing a positive weight $x_a$ to each $a$-particle.
We have
\beq
Z(z,x_1,\dots,x_A)=\frad{1}{1-\sum_{a=1}^A\frac{I_a(x_az)}{1+I_a(x_az)}}.
\label{zares}
\eeq

The expressions~(\ref{nares}) and~(\ref{zares}) hold in full generality.
In the case where all generating series $I_a(z)$ are rational,
the resulting series $\N(z)$ and $Z(z,x_1,\dots,x_A)$
are also rational functions of $z$ and of the weights $x_a$.

The simplest ensemble in the rational class is again the flat one,
where there are no constraints at all on the cluster lengths.
This generalizes the ensemble considered in section~\ref{flat}.
We have
\beq
I_a(z)=\frac{z}{1-z}
\eeq
for all species $a$.
The formula~(\ref{nares}) reads
\beq
\N(z)=\frac{1}{1-Az},
\eeq
and so
\beq
\N_N=A^N.
\eeq
The formula~(\ref{zares}) reads
\beq
Z(z,x_1,\dots,x_A)=\frad{1}{1-z\sum_{a=1}^Ax_a},
\eeq
and so
\beq
\N_{N,M_1,\dots,M_A}
={N\choose M_1,\dots,M_A}=\frac{N!}{\displaystyle{\prod_{a=1}^AM_a!}}
\eeq
is the multinomial coefficient, with the constraint $M_1+\cdots+M_A=N$.
The above results express that each site is independently occupied by a particle
of any species $a=1,\dots,A$.

\section{Transfer-matrix formalism}
\label{tm}

In this appendix we demonstrate the equivalence between the renewal approach
put forward in this work and the transfer-matrix approach,
on the explicit example of the ensemble where empty sites are isolated,
investigated in section~\ref{isola}.

The key point of the transfer-matrix formalism consists
in introducing partial partition functions,
defined by assigning fixed values to the occupations of the few rightmost sites,
in such a way that these partition functions obey closed linear recursions.
In the present case,
we introduce $Z_N^\u(x)$ and~$Z_N^\z(x)$,
defined by conditioning the configurations on the state (occupied or empty)
of the rightmost site.
We have then $Z_N(x)=Z_N^\u(x)+Z_N^\z(x)$.

The constraint that empty sites are isolated
implies that the partial partition functions obey the recursion
\beq
\pmatrix{Z_{N+1}^\u(x)\cr Z_{N+1}^\z(x)}
=\T(x)
\pmatrix{Z_N^\u(x)\cr Z_N^\z(x)},
\label{trec}
\eeq
where $\T(x)$ is the $2\times2$ transfer matrix
\beq
\T(x)=
\pmatrix{x&x\cr 1&0}.
\eeq
The characteristic polynomial of $\T(x)$ reads
\beq
P(\lam,x)=\lam^2-x\lam-x,
\eeq
so that its eigenvalues are
\beq
\lam_\pm(x)=\frac{x\pm\sqrt{x(x+4)}}{2}.
\eeq
The largest eigenvalue of $\T(x)$ obeys
\beq
\lam_+(x)=\frac{1}{z_0(x)},
\label{lamiden}
\eeq
where $z_0(x)$ is the nearest root of the denominator $D(z,x)$ (see~(\ref{zasy})),
which enters the analysis of the thermodynamic limit.

The initial conditions to the recursion~(\ref{trec}) read $Z_1^\u(x)=x$ and $Z_1^\z(x)=1$,
as there is a single configuration of each type ($\u$~and~$\z$).
These conditions translate to $Z_0^\u(x)=1$ and $Z_0^\z(x)=0$.
The partial generating series $Z^\u(z,x)$ and $Z^\z(z,x)$,
defined in analogy with~(\ref{hzdef}),
are therefore given by
\beq
\pmatrix{Z^\u(z,x)\cr Z^\z(z,x)}
=(1-z\T(x))^{-1}
\pmatrix{1\cr0}.
\label{matinv}
\eeq
We have
\beq
(1-z\T(x))^{-1}=\frac{1}{1-xz-xz^2}\pmatrix{1&xz\cr z&1-xz},
\label{tinv}
\eeq
and so
\beq
Z^\u(z,x)=\frac{1}{1-xz-xz^2},\qquad
Z^\z(z,x)=\frac{z}{1-xz-xz^2},
\label{tz}
\eeq
and finally
\beq
Z(z,x)=\frac{1+z}{1-xz-xz^2},
\eeq
in agreement with~(\ref{z2}).

Setting $x=1$ in~(\ref{tz}), we obtain
\beq
\N^\u(z)=\frac{1}{1-z-z^2},\qquad
\N^\z(z)=\frac{z}{1-z-z^2},
\eeq
and so
\beq
\N_N^\u=F_{N+1},\qquad
\N_N^\z=F_N,
\eeq
where $F_n$ are the Fibonacci numbers (see~(\ref{fibo})), and finally
\beq
\N_N=F_{N+2},
\eeq
in agreement with~(\ref{n2}).

The expression~(\ref{matinv}) implies that the denominator $D(z,x)$ entering~(\ref{tinv})
and~(\ref{tz}) is proportional to the characteristic polynomial $P(\lam,x)$ for $\lam=1/z$.
This property is fully general.
It implies that the degree~$\Delta$ of $D(z,x)$ coincides
with the dimension of the transfer matrix $\T(x)$.
It also yields the identity~(\ref{lamiden}).

\section*{References}

\bibliography{paper.bib}

\end{document}